\documentclass[aps,prl,twocolumn,superscriptaddress,tightenlines,longbibliography,reprint]{revtex4-1}
\usepackage{graphicx, color}
\usepackage{amsmath, amssymb, amsfonts, mathrsfs, ulem, comment}
\usepackage{times}
\usepackage[margin=1in]{geometry}

\begin{document}

\title{Designing Surface Charge Patterns for Shape Control of Deformable Nanoparticles}
\author{Nicholas E. Brunk}
\affiliation{Intelligent Systems Engineering, Indiana University, Bloomington, Indiana 47408, USA}
\author{JCS Kadupitiya}
\affiliation{Intelligent Systems Engineering, Indiana University, Bloomington, Indiana 47408, USA}
\author{Vikram Jadhao}
\email{vjadhao@iu.edu}
\affiliation{Intelligent Systems Engineering, Indiana University, Bloomington, Indiana 47408, USA}

\begin{abstract}
    Designing reconfigurable materials based on deformable nanoparticles (NPs) hinges on an understanding of the energetically-favored shapes these NPs can adopt. Using simulations, we show that hollow, deformable patchy NPs tailored with surface charge patterns such as Janus patches, stripes, and polyhedrally-distributed patches differently adapt their shape in response to changes in patterns and ionic strength, transforming into capsules, hemispheres, variably-dimpled bowls, and polyhedra. The links between anisotropy in NP surface charge, shape, and the elastic energy density are discussed.
\end{abstract}

\maketitle

Advances in nanotechnology have revolutionized the capacity to fabricate nanoparticles (NPs) with exquisite control of their surface properties \cite{mitragotri2007RodsDiscsStretched, klinger2014RodPolymersSurfactants, van2015SynthesisIPC}.
The broad design space spanned by surface modification and NP shape has enabled the fabrication of application-specific superstructures via various assembly engineering strategies \cite{akcora2009anisotropic,kegel2012DumbbellPatchyAssembly, glotzer2015, rossi2015shape,reddy2018stable}, including the use of charge-patterned NPs and electrostatic control \cite{andelman2017PatchySurfacesReview, chen2017JanusAssembly, velev2010PartialJanusAssembly, monica2016PartialJanusAssembly,gao2019electrostatic}.
Inspired by biology, there is a keen interest in how self-assembly is affected by building blocks that dynamically respond to external stimuli and reconfigure during assembly \cite{nguyen2011self,batista2010crystallization,gang2011shaping,zhang2011continuous,bian2011shape}. This dynamic dimension enables unique structural organizations via otherwise inaccessible assembly pathways, broadening assembly engineering approaches to design reconfigurable materials \cite{nguyen2011self,gang2011shaping}.  
For example, computational studies showed that shape-flexible nanorods can overcome kinetic barriers to yield reconfigurable structures that switch between various assembled mesophases such as square grids and bilayer sheets \cite{nguyen2010reconfigurable,nguyen2011self}. 
Phase transformations from simple cubic to rhombohedral structures were realized experimentally by accommodating the shape evolution of NPs from cubes to spheres \cite{zhang2011continuous}. 

Shape-changing NPs have also received attention as candidates for designing stimuli-responsive nanocontainers in therapeutic applications \cite{abadeer2016Review,liu2012shape,zhou2011pHTargeting}, where studies have demonstrated that the cellular uptake of nanocarriers is affected by their shape, size, charge, and deformability \cite{higuchi2019UptakeSims, chithrani2010UptakeTransport,nangia2012effects,liu2012shape}. 
Many techniques have been developed to synthesize such nanostructures \cite{blum2015stimuli,yoo2010EnvShapeSwitch, klinger2014RodPolymersSurfactants,williford2014ShapePolyDNA} utilizing different materials including biomolecular constructs and polymer-based vesicles where electrostatic interactions play a critical role in inducing shape changes via the control of pH or ionic strength \cite{blum2015stimuli,checot2003supramolecular,gebhardt2007rod,yoo2010EnvShapeSwitch}.

Here, we address a class of dynamic nanoscale building blocks: deformable NPs whose surface is tailored with charged patches. Although utilization of rigid, patchy NPs is common in assembling desired structures, little is known regarding the utilization of spontaneous NP deformation based on the location, size, or number of surface patches. 
Computational studies of deformations of patchy, flexible NPs have been largely limited to the use of elastic surface inhomogeneities, yielding, for example, regular polyhedra, buckled conformations, and collapsed bowl-like structures \cite{monica2011MulticomponentMembranes, nelson2013BucklingWeakSpots,  monica2012ShapesSoftBoundaries}. 
Alternatively, theoretical efforts have focused on revealing shape transitions in homogeneously-charged flexible nanostructures as a function of ionic conditions \cite{jadhao2014OriginalShapes, jadhao2015ShapesCoulomb, brunk2019ShapesTension, yao2016electrostatics} and understanding shape manipulation in uncharged elastic shells driven by topological defects, compression, or magnetic forces \cite{nelson2003VirusBuckling,  gompper2011CompressionDeformation, monica2019MagneticShapes}. Despite emerging capabilities for synthesizing charge-patterned NPs and colloids \cite{van2015SynthesisIPC, chang2019JanusDefinition, yi2016JanusDefinition} and the prevalent use of electrostatic control to induce changes in material assembly behavior \cite{gao2019electrostatic,brunk2019VLPAssembly}, a thorough understanding of the interplay between electrostatic and elastic energies for shape manipulation in charge-patterned flexible NPs is lacking \cite{monica2011IonicBuckling}.

We study a variety of shape-switching scenarios using a continuum model of deformable, hollow NPs to elucidate the relationship between surface charge patterns, deformability, and the low-energy shapes of NPs. 
We show that surface charge patterns can induce deformations in flexible NPs for a broad range of salt concentration. 
Our simulations reveal that NPs with charged stripes, Janus patches, and polyhedrally-distributed patches differently adapt their shape in response to changes in salt concentration over $0.05 - 20$ mM, transforming into rod-like structures, capsules, spinning-tops, hemispheres, variably dimpled bowls, and particles with polyhedral protrusions.
These deformations controlled by pattern type and screening length can change the directional specificity of interactions between NPs, which may in turn hinder or promote reconfiguration in assembled materials. 
The charge-pattern-based shape control of deformable NPs revealed in this work has implications for the design of responsive nanocontainers and dynamic building blocks for assembling reconfigurable materials.  

The model NPs are initialized as spherical hollow containers of radius $R = 20$ nm. Bending modulus $\kappa_b$ and stretching constant $\kappa_s = k_sR^2$, where $k_s$ is the spring constant (proportional to the Young's modulus), are introduced to effectively account for the intermolecular short-range interactions that resist deformation and preserve the structural integrity of the NP. These elastic parameters are chosen to be in a regime where the equilibrium shape without any surface charge is a sphere. Spontaneous buckling is suppressed by choosing a low Foppl von Karman (FvK) number \cite{nelson2003VirusBuckling}. 
Long-range electrostatic interactions that drive the shape deformation are included via a screened Coulomb potential with screening length $\lambda$ representing the effects of mobile ions in solutions inhabiting the NP. The surface charge patterns are restricted to designed patches exhibiting charges of only one sign. 
The NP surface is discretized with $N_v$ vertices, constituting $N_e$ edges and $N_f$ triangular faces. Using the discretization of the continuum expression for the elastic energy \cite{nelson1}, the Hamiltonian $\mathcal{H}$ characterizing the NP system \cite{jadhao2014OriginalShapes,brunk2019ShapesTension} is written in units of $k_BT$ as
\begin{align}\label{eq:mesh_energy}
    \mathcal{H} &= 
    \frac{\kappa_b}{2} \sum_{l = 1}^{N_e} |\vec{n}_{l_1} - \vec{n}_{l_2}|^2 
    + \frac{\kappa_s}{2 R^2} \sum_{l = 1}^{N_e} (|\vec{r}_{l_1} - \vec{r}_{l_2}| - a_l)^2 \nonumber\\ 
	&+ \frac{l_B}{2} \sum_{i = 1}^{i=N_v} \sum_{j\ne i}^{j = N_v} q_i q_j e^{-|\vec{r}_i - \vec{r}_j|/\lambda} / |\vec{r}_i - \vec{r}_j|,
\end{align}
where the first, second, and third term are the total bending, stretching, and electrostatic energy respectively.
$\mathcal{H}$ is extended to include a volume constraint term which allows NP volume changes on the order of $10\%$ for shape deformations shown here 
\footnote{See Supplemental Material for details of the model, simulation methods, shape descriptors, and analysis of ion condensation effects}.
In Equation \ref{eq:mesh_energy}, $\vec{n}$ denotes the normal vector associated with a face, $a_l$ is the equilibrium length of edge $l$, and $l_B$ is Bjerrum length in water. $\vec{r}_i$ denotes the position vector of vertex $i$ that parameterizes the NP shape and $q_i$ is the charge associated with this vertex, determined by the charge pattern. 

\begin{figure}[ht]
\centerline{\includegraphics[scale=0.14]{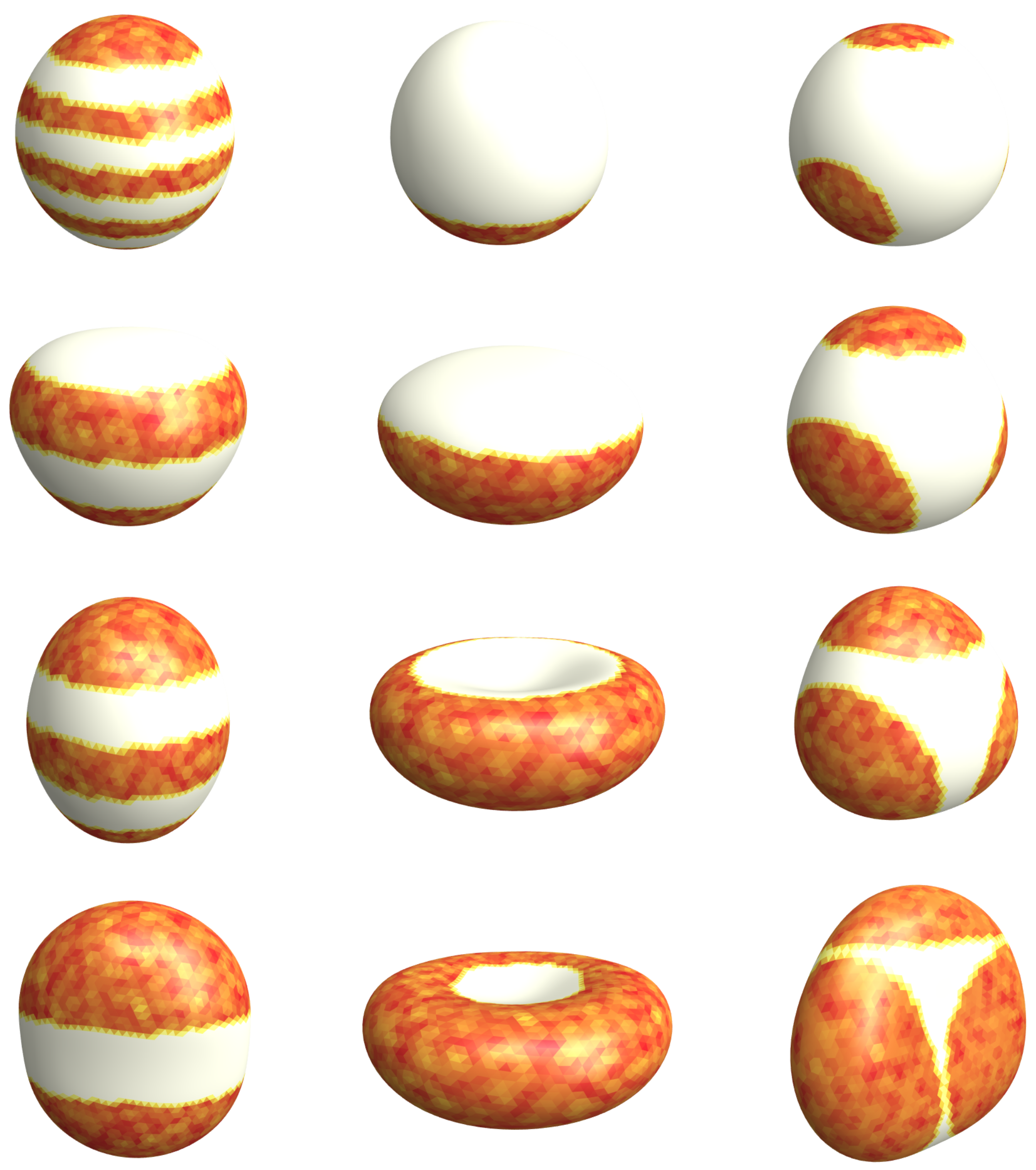}}
\caption
{\label{PatternControl}
Snapshots of low-energy shapes of charge-patterned flexible nanoparticles (NPs) for striped patterns (left column), Janus patches (middle column), and tetrahedrally-distributed patches (right column). 
All NPs have initial volume $4/3 \pi 20^3 \mathrm{nm}^3$, with charged regions (orange) of charge density $0.12 e/\mathrm{nm}^2$ in a salt solution at concentration $1$ mM. 
Shifting the pattern (e.g., by changing the fractional charge coverage) leads to variations in the NP shape. Striped NPs form spherocylinders with rounded or flattened tops, Janus patches produce hemispheres and flattened bowls, and tetrahedrally-distributed patches produce variably-rounded tetrahedra.
}
\end{figure}

To obtain the low-energy shapes of the NP, $\mathcal{H}$ is minimized subject to the volume constraint using a molecular dynamics based simulated annealing procedure  \cite{jadhao2014OriginalShapes,brunk2019ShapesTension} accelerated with an OpenMP/MPI hybrid parallelization technique \footnote{Code available at https://github.com/softmaterialslab/np-shape-lab/}.
The low-energy NP shapes are controlled by the elastic moduli, the charge pattern, and the screening length. Note that explicit counterion condensation may lead to a reduction in the effective charge associated with the NP. The effects of this charge renormalization are assessed by performing molecular dynamics simulations of counterions near a NP with a shape produced via the energy minimization process \cite{Note1}.

\begin{figure}
\centerline{\includegraphics[scale=0.09]{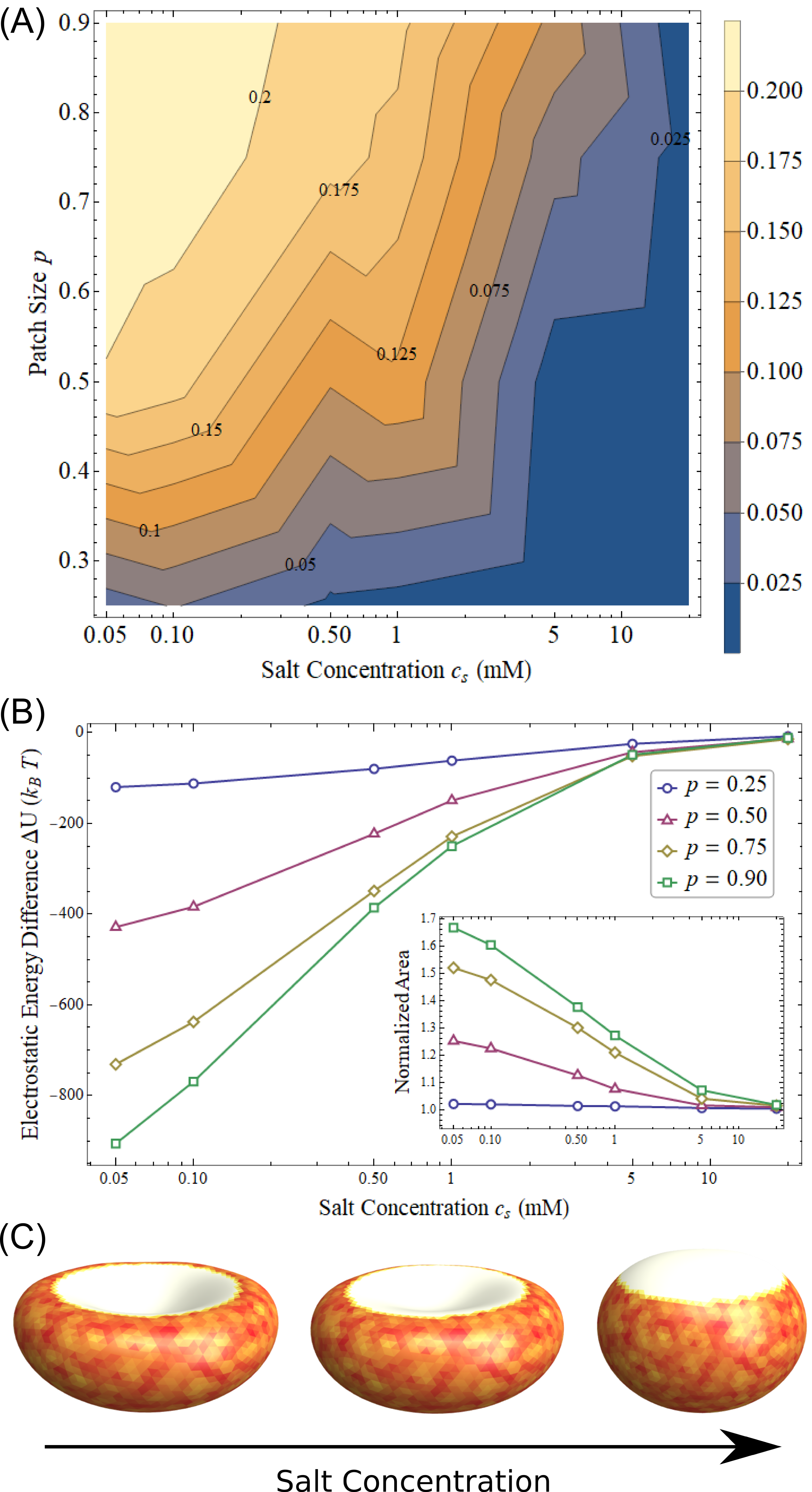}}
\caption
{\label{Janus_SaltControl}
Pattern-shape links analyzed for nanoparticles (NPs) patterned with Janus patches with fractional surface charge coverage $p \in (0.25, 0.9)$ (middle column in Figure \ref{PatternControl}) for varying salt concentration $c_s \in (0.05, 20)$ mM. 
(A) Asphericity as a function of $c_s$ and $p$. NPs associated with large $p$ and small $c_s$ exhibit higher asphericity, signaling greater deviations from the spherical conformation.
(B) Changes in NP electrostatic energy and area (inset) vs. $c_s$ for different $p$. Shape transitions are accompanied with a decrease in the electrostatic energy and an increase in the area relative to the spherical conformation.
(C) Representative snapshots of the low-energy shapes of Janus NPs with $p = 0.75$ as $c_s$ increases from left to right. 
}
\end{figure}

Figure \ref{PatternControl} shows the low-energy conformations of NPs for three distinct sets of designed surface charge patterns: alternating charged and neutral stripes of equal area, Janus patches with varying fractional surface charge coverage, and tetrahedrally-distributed patches with varying patch size. For each set, the different patches have the same initial local charge density $\sigma = 0.12 e/{\mathrm{nm}}^2$.
All NPs are in an aqueous solution with salt concentration $c_s = 1$ mM and have elastic moduli of $\kappa_b = 5$ and $\kappa_s = 125$ (in units $k_B T$).
NPs having striped patches yield spherocylinders including rod-like and capsule-shaped structures as the number of charged stripes is changed. Shapes with rounded or flattened ends are observed depending on the charged or neutral nature of the terminal ends. 
For Janus patches with fractional surface charge coverage $p$, 
the NP exhibits shape transitions from sphere ($p=0$, fully uncharged) to hemisphere ($p=0.5$) to variably-dimpled bowls ($p=0.75, 0.9$).
NPs with charged patches centered around the sites of circumscribed tetrahedral vertices expand to generate variably rounded tetrahedral-shaped NPs, depending upon the patch size. 
For each of the three distinct types of charge patterns, as the uniformly-charged limit is approached (e.g., $p=1$ for Janus patches), the low-energy shape converges to a disc conformation observed previously for uniformly-charged deformable NPs \cite{jadhao2014OriginalShapes,jadhao2015ShapesCoulomb,brunk2019ShapesTension}.

For a given patterned NP, its shape can be further controlled by modifying solution conditions such as salt concentration $c_s$.
The links between surface patterns and shapes associated with NPs under varying $c_s$ can be presented in the form of 2-dimensional contour plots (maps) using shape metrics such as asphericity. Figure \ref{Janus_SaltControl}A shows an asphericity shape map for NPs with Janus patches as a function of $c_s$ and fractional surface charge coverage $p$ \cite{Note1}. 
The simulated solution conditions and NP elastic properties are otherwise the same as in Figure \ref{PatternControl}.
The map is constructed via an interpolation of the data for the low-energy shapes generated from a set of 24 simulations of deformable NPs with $p = 0.25, 0.5, 0.75, 0.9$ and $c_s = 0.05, 0.1, 0.5, 1, 5, 20$ mM \footnote{Snapshots of 24 nanoparticle shapes are shown in the Supplemental Material.}. Asphericity contours demarcate regions of the design space capable of deforming NPs to a given extent, with deformation determined by $p$ and $c_s$. 
Other shape descriptors such as relative anisotropy yield similar results \cite{Note1}.

\begin{figure}
\centerline{\includegraphics[scale=0.18]{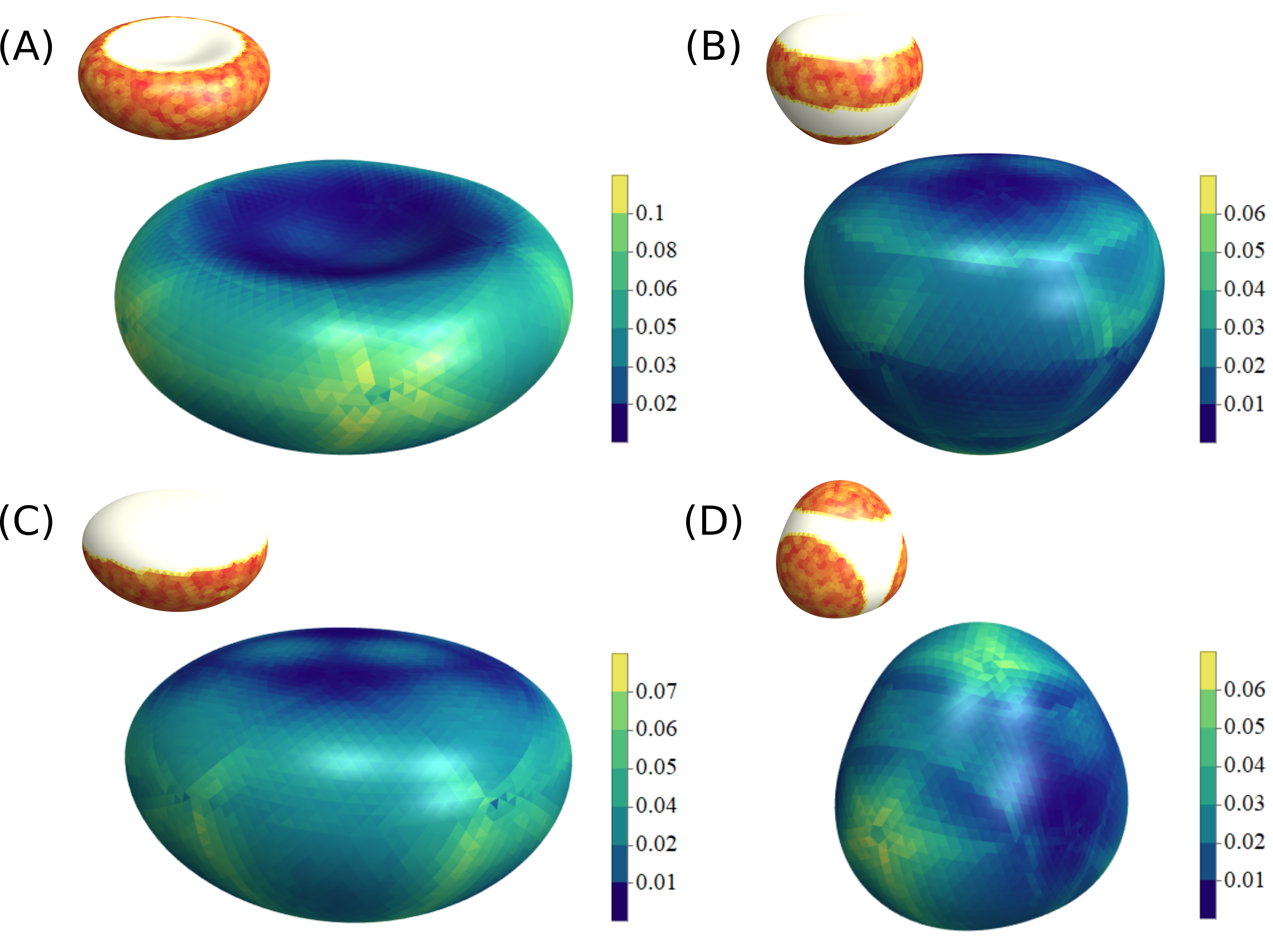}}
\caption
{\label{elastic_energy}
The elastic energy distribution on the surface of the deformed nanoparticle (NP) for a representative set of NPs tailored with surface charge patterns shown in the top-left. 
NPs with Janus patches characterized by fractional charge coverage $p = 0.75$ (A) and $p = 0.5$ (C) exhibit a ``Janus-like'' elastic energy density. A similar coupling between anisotropy in charge pattern and elastic energy density is observed in striped NPs (B) and NPs with polyhedrally-distributed patches (D).  
}
\end{figure}

Figure \ref{Janus_SaltControl}B shows the changes in the total electrostatic energy and the area of the NP as a function of $c_s$ for different Janus patches characterized with $p=0.25,0.5,0.75,0.9$. 
In all cases, the difference $\Delta U$ between the electrostatic energy of the deformed NP and the undeformed (initial) NP is negative, indicating that the decrease in electrostatic energy drives the shape deformation. The NP lowers its electrostatic energy at the cost of an increase in its elastic energy and area. Similar trends are observed for NPs having striped patterns and polyhedrally-distributed patches.
Figure \ref{Janus_SaltControl}C provides snapshots of a specific shape transition (from dimpled to flattened bowls) recorded in Figures \ref{Janus_SaltControl}A and B for $p = 0.75$ as $c_s$ is increased from $0.5$ to $5$ mM. 

\begin{figure*}[htb]
\centerline{\includegraphics[scale=0.4]{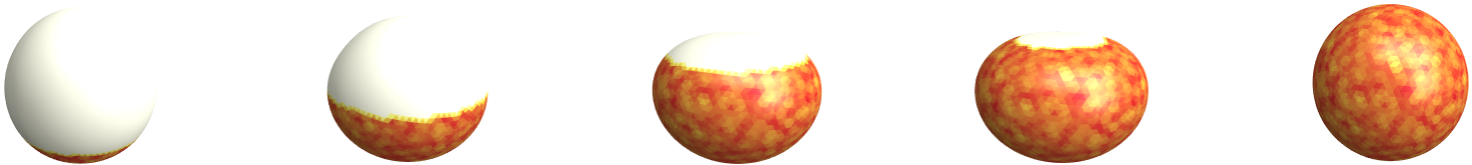}}
\caption
{\label{PatternImportance}
Low-energy shapes of NPs with Janus patches of charge density $\sigma = 0.06 e/\mathrm{nm}^2$ and elastic moduli $\kappa_b = 3$ and $\kappa_s = 75$ (in units of $k_BT$). Patch size increases from left to right as $p = {0.25, 0.50, 0.75, 0.90, 1.0}$; other parameters are the same as in Figure \ref{PatternControl}. Anisotropy of the surface charge itself causes deformation; when uniformly-charged ($p=1$), the NP does not deform.
}
\end{figure*}

Figure \ref{elastic_energy} shows the distribution of local elastic energy density (sum of the bending and stretching energies) associated with a representative set of deformed NPs. The inhomogeneous elastic energy density  exhibits a coupling to the anisotropic surface charge distribution in each case. For example, NPs with Janus patches exhibit a ``Janus-like'' elastic energy density, where an expansion of charged regions of the NP results in a higher local elastic energy compared to the uncharged regions. 

Charge anisotropy induces shape deformation, even for the case of low charge density $\sigma$ where NPs do not deform when the surface charge is uniformly distributed. We examined the shapes of NPs with Janus patches for $\sigma = 0.06 e/\mathrm{nm}^2$ and elastic moduli approximately half of the values employed in NPs investigated in Figure \ref{PatternControl}. When uniformly-charged, the NP does not deform at $c_s = 1$ mM. However, generating an inhomogeneity in the form of an uncharged region induces deformation despite lowering the net surface charge (Figure \ref{PatternImportance}). Tuning the anisotropy of the surface charge deforms the NP into egg-like, flattened bowl, and spinning-top shapes.

The charge density $\sigma$ in the NP model implicitly includes the effects of charge renormalization due to ion condensation. All charged patches in Figure \ref{PatternControl} have the same $\sigma = 0.12e/\mathrm{nm}^2$. The NP surface charge, however, varies with the pattern and is largest ($\approx 600e$) for the homogeneously-charged NP. Molecular dynamics (MD) simulations of monovalent counterions surrounding undeformed (spherical) NPs with charge patterns depicted in Figure \ref{PatternControl} show that patchy NPs experience less ion condensation compared to homogeneously-charged NPs.
For example, the fraction $\alpha$ of condensed ions for NPs with Janus patches monotonically increases with increasing patch size \cite{Note1}.
As condensation effects are most severe for the uniformly-charged spherical NP, the latter can be used to assess the feasibility of realizing a charge density of $\sigma = 0.12e/\mathrm{nm}^2$ in the event of ion condensation, which is known to limit the maximum effective charge or the critical valence \cite{pincus2015ShapesCondensation}. 
The extent to which ion condensation lowers the critical valence is dependent on NP packing fraction $\eta$ and salt concentration $c_s$. 
Mean-field calculations based on the Manning two-state model \cite{manning2007counterion,jadhao2014OriginalShapes, jadhao2015ShapesCoulomb, brunk2019ShapesTension, Note1} show that uniformly-charged spherical NPs under salt-free conditions have a critical valence greater than $\sigma = 0.12 e/\mathrm{nm}^2$ for $\eta < 10^{-8}$. By the above arguments, this $\sigma$ value is also feasible for charge-patterned NPs for similarly low $\eta$. 

For high NP packing fraction, $\sigma$ value in the coarse-grained model needs to be lowered as stronger condensation effects are expected  \cite{brunk2019ShapesTension,pincus2015ShapesCondensation}. MD simulations of counterions surrounding charge-patterned NPs characterized with an initial bare charge density $\sigma_b = 0.12 e/\mathrm{nm}^2$, show significant condensation for both undeformed and deformed NPs (Figure \ref{PatternControl}) at $\eta = 10^{-2}$ \cite{Note1}. Associated lowering of the effective charge density $\sigma$ and the related reduction in the electrostatic drive to deform can be counteracted by reducing the elastic inhibition to deformation by synthesizing less rigid NPs. 
For example, deformations similar to those found for NPs with Janus patches in Figure \ref{PatternControl} (middle column) are achieved with lower $\sigma = 0.09e/\mathrm{nm}^2$ for NPs with reduced elastic parameters: $\kappa_b = 3$ and $\kappa_s = 75$ \cite{Note1}. 

We also note that the difference in the extent of ion liberation between spherical and deformed patterned NPs is small for all patterns, as measured by the difference in the respective condensate fractions extracted via MD simulations. For example, in the case of NPs with Janus patches, this value is within a factor of 2 of typical statistical fluctuations in $\alpha$ for deformed and undeformed NPs \cite{Note1}.
We thus do not expect ion release and associated entropic effects to sufficiently counteract the electrostatic drive to deform and stabilize the spherical conformation as has been found for highly-charged nanocontainers \cite{pincus2015ShapesCondensation}. 
However, the effects of explicit counterion-NP interactions during the shape change process are neglected in our model, and may inhibit deformation beyond charge renormalization.  
Further, assumptions of static (quenched) charge patterns may not hold in experimental conditions where the timescales of dissociation of charged moieties are comparable to those of shape deformation. Future work is needed to address these model limitations as well as the role of kinetic effects.

We have demonstrated the capacity to electrostatically control the shape of hollow, deformable NPs via the synthesis of specific surface charge patterns.
Our simulations showed that NPs with Janus patches, stripes, and polyhedrally-distributed patches adapt their shape in response to pattern tuning, transforming into different structures including spherocylinders, spinning-tops, hemispheres, bowls, and polyhedral particles. The shape transitions can be controlled by tuning salt concentration. The model can be readily extended to address other types of charge patterns, including oppositely-charged patches, as well as to probe patterned NPs of different sizes \cite{brunk2019ShapesTension}. 
Our findings have implications for the design of shape-tunable nanoscale building blocks for assembling charged polymer-based reconfigurable materials. 

This work is supported by the NSF through awards DMR-1753182 and 1720625. Simulations were performed using the Big Red supercomputing systems.

%\bibliography{References}

%

\mbox{~}
\clearpage

\section{\underline{\large{Supplemental Material}}}

\section{Details of the Simulation Methods}
\paragraph{Molecular Dynamics based Simulated Annealing}
In the coarse-grained model representation of the hollow, deformable nanoparticle (NP), the Hamiltonian $\mathcal{H}$ (given by Equation 1 of the main text) is a function of the position vectors of the vertices discretizing the NP surface. These vectors parameterize the shape of the NP. The low-energy shapes correspond to the minimum of $\mathcal{H}$ subject to the volume constraint. 
The volume constraint is imposed as a quadratic term: $\lambda_v \sum_{k \in N_f} (V_k - V_{k_0})^2$, where $V_k$ is the volume subtended by the $k^{\mathrm{th}}$ face of the triangulated surface, $V_{k_0}$ is its initial volume, $N_f$ is the total number of faces, and  $\lambda_v$ denotes the strength of the constraint. The total initial volume of the NP is $4/3\pi R^3$, where $R$ is the radius of the initial, pre-assembled spherical NP.
$\lambda_v$ is chosen such that it allows variation in the volume generally up to $10\%$. 

The constrained energy minimization is carried out using a molecular dynamics based simulated annealing procedure described in our earlier work \cite{jadhao2014OriginalShapes, brunk2019ShapesTension}. Our code is publicly available on GitHub \footnote{Code available at https://github.com/softmaterialslab/np-shape-lab/. Associated simulation tool available online at https://nanohub.org/tools/npshapelab}. Simulation setup, analysis, and visualization were conducted in Mathematica 12.1 \cite{Mathematica}.
The simulation begins at a fictitious temperature high enough to allow the NP surface to deform by exploring the energy landscape, after which kinetic energy is gradually removed from the system regulated via a Nose-Hoover thermostat. This allows a gradual convergence upon the fluctuating shapes of low energy. We note that the NP shape can get trapped in some metastable state in simulations. As a result, it is likely that the low-energy or ``equilibrium'' shapes found in simulations are not the true ground states of the model Hamiltonian. The low-energy conformations can be considered as modes that describe the characteristic behavior of the NP as it approaches the true ground state. 

Simulations are accelerated using an OpenMP/MPI hybrid parallelization technique that generates a speedup of up to $20 \times$ (using 16 nodes and 24 processors per node).
This acceleration enabled over $7000$ simulations of NPs discretized with mesh points up to $\approx 4000$, expediting the exploration of equilibrium NP shapes for a broad range of surface patterns and solution conditions.

\paragraph{Molecular Dynamics Simulations of Ions near NP surface}
To assess the effects of ion condensation on the patterned NP surface under salt-free conditions, molecular dynamics simulations of ions near NPs are performed using LAMMPS \cite{PlimptonLAMMPS}.
We briefly describe the key implementation steps below; additional details are provided in an earlier paper \cite{brunk2019ShapesTension}.
Monovalent counterions are placed in a periodic cubic box enclosing a NP with a shape produced via the energy minimization process. The number of counterions is selected to neutralize the charge on the NP and vary between 150 and 600, depending on the surface charge pattern.
The volume of the box is determined by the desired NP packing fraction. The NP surface is finely meshed with 4692 to 5592 mesh points, depending on the surface area, with each point representing a counterion-sized spherical bead such that the surface is impermeable to ions. 
Ions are modeled as spherical particles of diameter $d = 0.6$ nm. Ion-ion and ion-NP steric interactions are modeled using the standard purely-repulsive Lennard-Jones potential with a cutoff of $2^{1/6}d$. Ion-ion and ion-NP electrostatic interactions are modeled using the Coulomb potential. Long-range effects are treated using the PPPM method.

Simulations are performed in an NVT ensemble at $298$ K. Ions and NP are assumed to be present in water modeled as a dielectric continuum (permittivity $\epsilon _r = 80$). 
On reaching equilibrium, Diehl's method is employed to extract the fraction $\alpha$ of ions condensed on the NP surface \cite{diehl2004effective}. 
The number of condensed ions is computed by comparing the electrostatic energy of ions binding to the NP surface ($U_{i}$) and the kinetic energy with which they might escape ($K_{i}$):  if $U_{i} \geq \chi K_{i}$, the $i^{\mathrm{th}}$ ion is considered condensed, where typically $\chi = 4/3$ \cite{diehl2004effective}.  

\section{Quantifying Shape Transitions}
The gyration tensor $\mathbf{S}$ is used to define the metrics quantifying the spatial distribution of the vertices describing the NP shape \cite{arkin2013ShapeMetrics}. 
$\mathbf{S}$ is a symmetric $3\times 3$ square matrix with diagonal components: $1/N_v\sum_{i=1}^{N_v} (x_i - \Bar{x})^2$, $1/N_v\sum_{i=1}^{N_v} (y_i - \Bar{y})^2$, $1/N_v\sum_{i=1}^{N_v} (z_i - \Bar{z})^2$ and off-diagonal components: $1/N_v\sum_{i=1}^{N_v} (x_i - \Bar{x}) (y_i - \Bar{y})$, $1/N_v\sum_{i=1}^{N_v} (x_i - \Bar{x}) (z_i - \Bar{z})$, $1/N_v\sum_{i=1}^{N_v} (y_i - \Bar{y}) (z_i - \Bar{z})$. Here, $(x_i,y_i,z_i)$ is the Cartesian coordinate of the position vector of the $i^{\mathrm{th}}$ vertex, and $N_v$ is the total number of vertices discretizing the NP surface. $(\Bar{x},\Bar{y},\Bar{z})$ is the Cartesian coordinate of the center of mass (e.g., $\Bar{x} = 1/N_v \sum_{i=1}^{N_v} x_i$). 

\begin{figure}[htb]
\centerline{\includegraphics[scale=0.32]{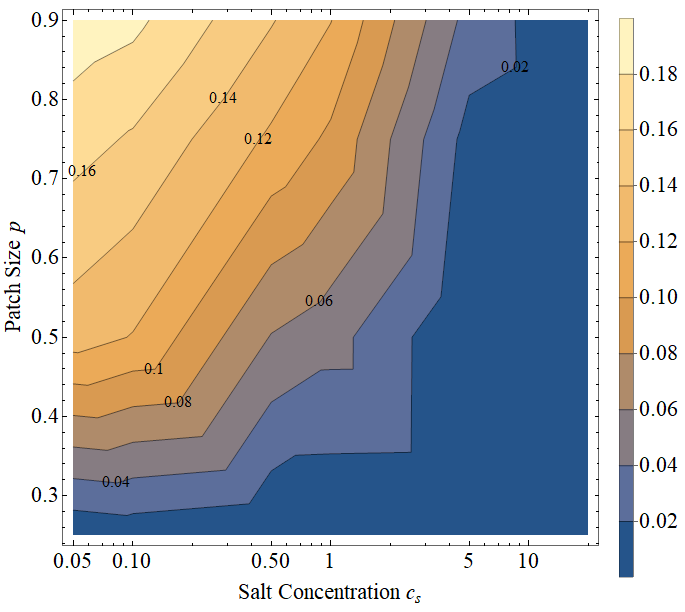}}
\caption
{\label{SI_ShapeMaps}
A map showing shapes of nanoparticles (NPs) patterned with Janus patches as a function of the fractional surface charge coverage or patch size $p \in (0.25, 0.9)$ and salt concentration $c_s \in (0.05, 20)$ mM. The shape map is generated using relative anisotropy as the shape descriptor. NPs associated with large $p$ and small $c_s$ exhibit higher relative anisotropy, signaling greater deviations from the spherical conformation. This map complements the map shown in Figure 2A of the main text that uses asphericity as the shape descriptor.
}
\end{figure}
The eigenvalues (principal moments) $\lambda_x^2, \lambda_y^2, \lambda_z^2$ of the gyration tensor can be used to define shape descriptors. We employ normalized asphericity $b^* = b / R_g^2$ as a shape descriptor, where $b$ is given by
\begin{equation}
\label{eq:Asphericity}
    b = \frac{3}{2} \lambda_z^2 - \frac{R_g^2}{2}, 
\end{equation}
$R_g^2 = \lambda_x^2 + \lambda_y^2 + \lambda_z^2$ is the squared radius of gyration, and axes are chosen such that $\lambda_x^2 \le \lambda_y^2 \le \lambda_z^2$. The normalization by $R_g^2$ facilitates the comparison between shapes of NPs of different sizes. Note $0 \le b^* \le 1$, where $b^*=0$ describes a spherically symmetric conformation. Relative anisotropy $\kappa^2 = \left( b^2 + \left(3/4\right)\left(\lambda_y^2 - \lambda_x^2\right) \right) / R_g^4$ is also used to characterize NP shapes (Figure \ref{SI_ShapeMaps}). $\kappa^2$ is also bounded between 0 and 1. 

\section{Details of the Analysis of Ion Condensation Effects}

In the main text, we describe studies where the shape deformation of NPs occurs for salt concentrations up to $c_s \approx 20$ mM when charged regions are characterized with an effective charge density of $\sigma = 0.12 e/\mathrm{nm}^2$ and the bending and stretching penalties (in units of $k_BT$) are $\kappa_b = 5$ and $\kappa_s = 125$, respectively. 
$\sigma$ and the associated net surface charge in the coarse-grained NP model implicitly include the effects of charge renormalization due to ion condensation. In the event of explicit ion condensation, the effective charge may be lowered, stabilizing the spherical conformation. These effects can be analyzed via mean-field calculations and molecular dynamics (MD) simulations of ions near NPs \cite{jadhao2014OriginalShapes,jadhao2015ShapesCoulomb,brunk2019ShapesTension}. The details of this analysis are provided here.  

\begin{figure}[htb]
\centerline{\includegraphics[scale=0.3]{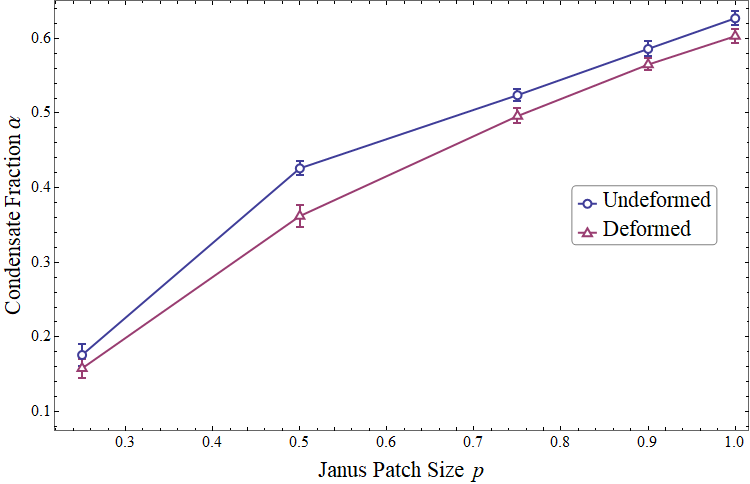}}
\caption
{\label{SI_alpha}
The ion condensate fraction $\alpha$ as a function of fractional surface charge coverage $p$ of NPs patterned with Janus patches. $\alpha$ rises monotonically with $p$ for both undeformed, spherical NPs (squares) and their corresponding deformed states (triangles) shown in Figure 1 of the main text. $\alpha$ is highest for the uniformly-charged case ($p = 1$). 
}
\end{figure}

We performed MD simulations of monovalent counterions surrounding undeformed (spherical) charge-patterned NPs characterized with an initial bare charge density of $0.12e/\mathrm{nm}^2$ and their corresponding deformed states shown in Figure \ref{PatternControl} of the main text. NP packing fraction was set to $\eta = 10^{-2}$ and ion trajectories were simulated in an NVT ensemble at 298 K under salt-free conditions. 
Regardless of the pattern type, we find ion condensation to be highest for NPs approaching the uniformly-charged limit. For example, in the case of NPs with Janus patches characterized with fractional surface charge coverage $p$, the condensate fraction $\alpha$ is highest for $p=1$ and decreases monotonically with $p$ (Figure \ref{SI_alpha}). 
Thus, charge renormalization effects are most severe for the uniformly-charged NP; as this case is also analytically tractable, mean-field calculations can be employed to analyze condensation effects.

We have shown previously via exact analytical calculation of the (unscreened) Coulomb energy of uniformly-charged and equipotential (conducting) spheroidal NPs that the spherical conformation is electrostatically unstable under the constraint of fixed volume against deformation to spheroidal structures (e.g., disc-like shapes formed as the fully-charged limit is approached in Figure 1 of the main text) \cite{jadhao2014OriginalShapes,jadhao2015ShapesCoulomb,brunk2019ShapesTension}. 
Using these results in the mean-field calculations based on the Manning two-state model \cite{manning2007counterion}, we showed that the uniformly-charged and equipotential spherical NPs with $\sigma$ up to $ 0.48e/\mathrm{nm}^2$ have a higher free energy relative to spheroidal morphologies in the event of ion condensation under salt-free conditions for a wide range of packing fraction $\eta \in (10^{-12},10^{-4})$ \cite{jadhao2014OriginalShapes, jadhao2015ShapesCoulomb, brunk2019ShapesTension}. Further, the maximum effective charge of a uniformly-charged spherical NP is found to be greater than $\sigma = 0.12 e/\mathrm{nm}^2$ for $\eta < 10^{-8}$, making the choice of $\sigma$ for the studies performed in the main text feasible for similarly low $\eta$.

\begin{figure*}[ht]
\centerline{\includegraphics[scale=0.4]{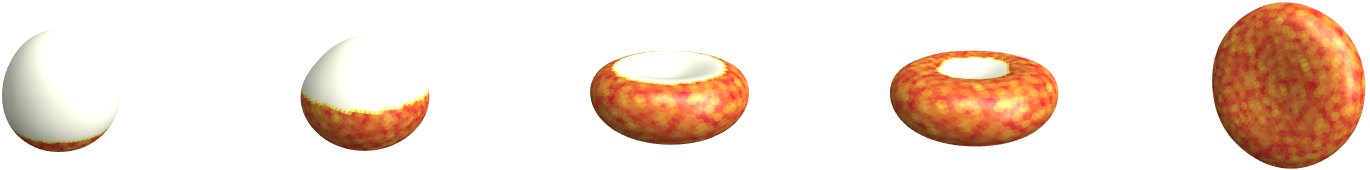}}
\caption
{\label{SI_LowerCharge_Control}
Deformations similar to those shown in Figure 1 (main) for NPs with Janus patches may be achieved with lower charge densities, e.g. $\sigma = 0.09 e/\mathrm{nm}^2$ shown here, provided the elastic moduli are reduced, e.g., to $\kappa_b = 3$ and $\kappa_s = 75$ (in units of $k_BT$ shown here. The size of the charged region increases from left to right as $p = 0.25, 0.50, 0.75, 0.90, 1.0$.
}
\end{figure*}

\begin{figure*}[bht]
\centerline{\includegraphics[scale=0.158]{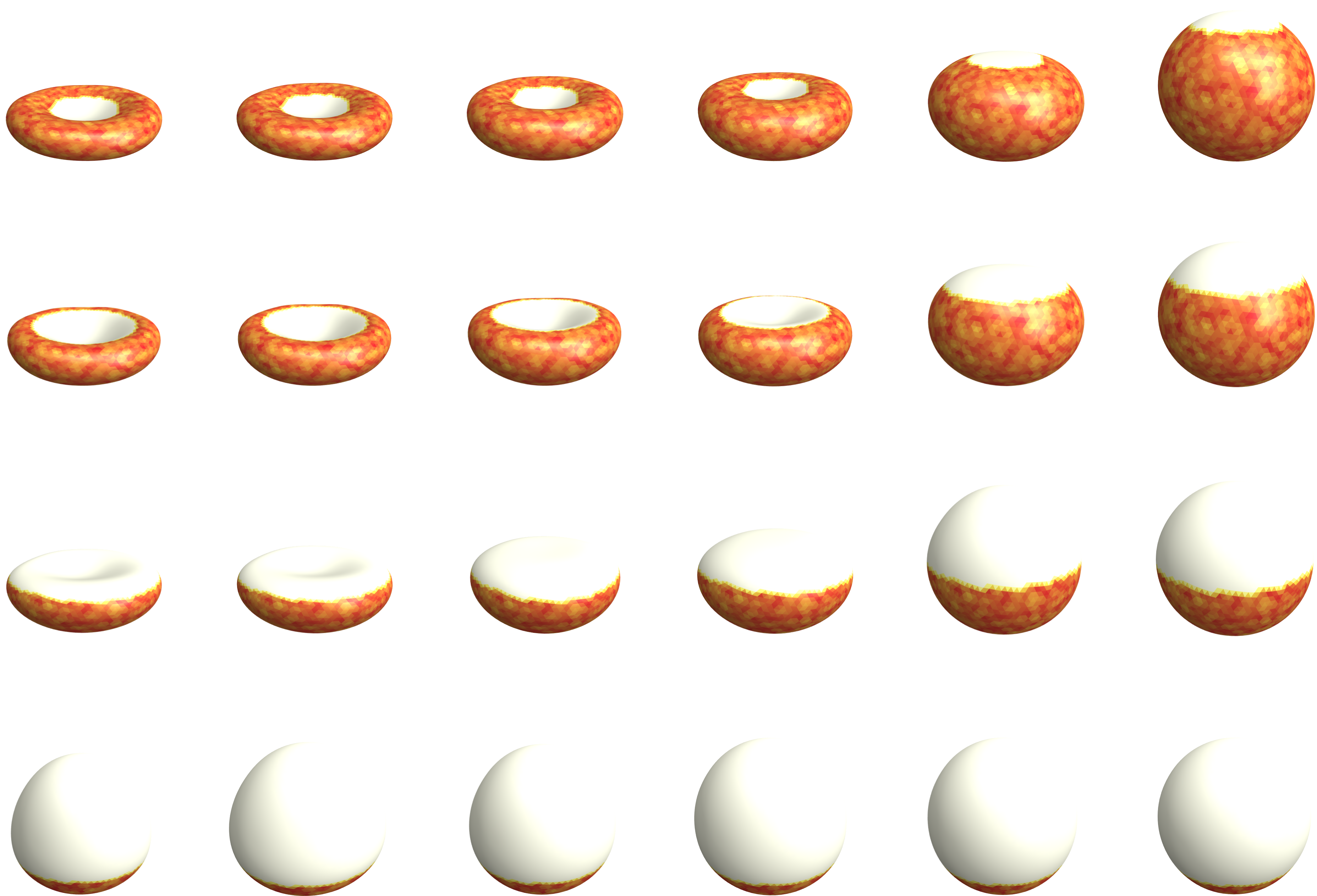}}
\caption
{\label{SI_gallery}
Snapshots of low-energy shapes of deformable NPs patterned with Janus charged patches for varying patch size $p$ (vertical axis) and salt concentration $c_s$ (horizontal axis). These shapes correspond to the dataset of 24 simulations utilized to generate the asphericity map in Figure 2A of the main text (and the anisotropy map in Figure 5 of the Supplemental Material), and they are associated with the results for the NP electrostatic energy and area presented in Figure 2B. As in Figure 2A, $p$ increases from bottom to top as $0.25, 0.50, 0.75, 0.90$, and $c_s$ increases from left to right as $c_s = 0.05, 0.1, 0.5, 1, 5, 20$ mM.} 
\end{figure*}

MD simulations show significant condensation for both undeformed and deformed NPs at a high packing fraction $\eta = 10^{-2}$. For example, in the study of NPs with Janus patches, ion condensate fraction can be as high as $\alpha \approx 0.6$ for charge patterns that exhibit significant surface coverage (Figure \ref{SI_alpha}).
We also find that regardless of the charge pattern, the difference in the extent of ion liberation between undeformed and deformed patterned NPs is small as measured by the difference in the respective condensate fractions $\Delta \alpha = \alpha_{s} - \alpha_{d}$. 
For example, the average value of this quantity for deformations of NPs with Janus patches shown in Figure \ref{PatternControl} (middle column) is $\Delta \alpha \approx -0.027$, which is within a factor of 2 of typical statistical fluctuations in $\alpha_s$ and $\alpha_d$ values (Figure \ref{SI_alpha}).
We thus do not expect ion release and associated entropic effects to sufficiently counteract the electrostatic drive to deform, as observed in simulations of ions near highly-charged spheroidal NPs \cite{pincus2015ShapesCondensation}. 
However, significant ion condensation suggests that $\sigma$ value in the coarse-grained model needs to be lowered for high $\eta$. 
We find that deformations similar to those seen for NPs with Janus patches in Figure \ref{PatternControl} (middle column) can be achieved for lower $\sigma$, e.g., $\sigma = 0.09e/\mathrm{nm}^2$, provided the elastic parameters are reduced (Figure \ref{SI_LowerCharge_Control}).

\end{document}